\documentclass[fleqn,usenatbib,useAMS]{mnras}
\usepackage{newtxtext,newtxmath}

\usepackage[T1]{fontenc}
\usepackage{ae,aecompl}

\usepackage{graphicx}	
\usepackage{amsmath}	
\usepackage{amssymb}	
\usepackage{wasysym}  
\usepackage{bm}       
\usepackage{etoolbox}
\usepackage{url}
\makeatletter 

  \patchcmd{\NAT@citex}
    {\@citea\NAT@hyper@{%
      \NAT@nmfmt{\NAT@nm}%
      \hyper@natlinkbreak{\NAT@aysep\NAT@spacechar}{\@citeb\@extra@b@citeb}%
      \NAT@date}}
    {\@citea\NAT@nmfmt{\NAT@nm}%
    \NAT@aysep\NAT@spacechar\NAT@hyper@{\NAT@date}}{}{}

  \patchcmd{\NAT@citex}
    {\@citea\NAT@hyper@{%
      \NAT@nmfmt{\NAT@nm}%
      \hyper@natlinkbreak{\NAT@spacechar\NAT@@open\if*#1*\else#1\NAT@spacechar\fi}%
        {\@citeb\@extra@b@citeb}%
      \NAT@date}}
    {\@citea\NAT@nmfmt{\NAT@nm}%
    \NAT@spacechar\NAT@@open\if*#1*\else#1\NAT@spacechar\fi\NAT@hyper@{\NAT@date}}
    {}{}
\makeatother

\DeclareMathOperator*{\proj}{proj}

\title[Blind nucleosynthetic source discovery in astronomical elemental abundance data]
{Blind nucleosynthetic source discovery in astronomical elemental abundance data}

\author[M.\ Milosavljevi\'c et al.]{
Milo\v s Milosavljevi\'c $^1$,
Patrick D.\ Aleo $^{1,2}$,
Natalie R.\ Hinkel $^{3,4}$, and
Haris Vikalo $^5$
\\
$^1$ Department of Astronomy, The University of Texas at Austin, Austin, TX 78712, USA \\
$^2$ Department of Astronomy, University of Illinois at Urbana-Champaign, Urbana, IL 61801, USA \\
$^3$ Department of Physics and Astronomy, Vanderbilt University, Nashville, TN 37235, USA \\
$^4$ Space Science and Engineering Division, Southwest Research Institute, San Antonio, TX 78238, USA \\
$^5$ Department of Electrical and Computer Engineering, The University of Texas at Austin, Austin, TX 78712, USA}

\date{Accepted XXX. Received YYY; in original form ZZZ}

\pubyear{2017}

\begin{document}

\maketitle

\begin{abstract}
We demonstrate physics-blind and purely data-based nucleosynthetic archetype discovery from heterogeneous stellar elemental abundance measurements. Archetype extraction is performed via non-negative matrix factorization (NMF). Because stellar elemental abundance measurements can be incomplete and sparsely sampled, we perform combined low-rank NMF and matrix completion on the Hypatia Catalog and JINAbase to extract archetypal elemental abundance patterns, the so-called \emph{endmembers}, and impute the missing measurements. We qualitatively compare the discovered nucleosynthetic archetypes to patterns anticipated from direct astrophysical nucleosynthetic calculations. Our tentative results indicate that the physics-blind analysis relying entirely on the incomplete chemical mixing in star formation can discern nucleosynthetic archetypes associated with asymptotic giant branch (AGB) stellar sources of the s-process, thermonuclear supernovae, and non-thermonuclear supernovae with an r-proces.
\end{abstract}

\begin{keywords}
methods: data analysis --- stars: abundances --- stars: AGB and post-AGB --- supernovae: general --- ISM: abundances --- ISM: evolution.
\end{keywords}

\section{Introduction}
\label{sec:introduction}

The Big Bang produced only the lightest atomic nuclei; of those, only hydrogen, helium, and lithium are sufficiently abundant to be astronomically detectable.  The naturally occurring nuclei beyond boron originate in stars, stellar explosions, and processes associated with stellar compact remnants: white dwarfs, neutron stars, and black holes. Nuclear reaction pathways synthesizing specific nuclides include the fusion of and capture of nucleons by nuclei, induced nuclear disintegration, and radioactive decays.  One or more representatives of these pathways can be realized in each of the various known nucleosynthetic contexts: nuclear fusion under hydrostatic conditions in stars, thermonuclear deflagrations and detonations within and on the surfaces of white dwarfs, turbulent nuclear burning behind stalled shocks following the collapse of massive stellar cores, sudden heating of stellar matter by strong shock waves, outflows from young neutron stars and from disrupted merging neutron stars, and outflows from the matter accreting onto black holes in the aftermath of neutron star mergers and black-hole-forming stellar core collapse. The isotopic yields of each pathway---the masses of isotopes synthesized---are functions of the initial composition (in particular the proton-to-nucleon ratio $Y_e$ and the related free neutron abundance) as well as the density, temperature or entropy, and in outflows and explosions---the time scale on which the material expands that determines which nuclei are ``frozen out of'' further collisional nuclear reactions.

Multiple pathways that occur under different conditions and on different time scales can inject their isotopic yields into the environment in the same galactic neighborhood.  Then their respective yields end up correlated in the elemental abundances of the new stars that form in the enriched galactic site.\footnote{For a review of stellar nucleosynthesis and galactic chemical enrichment see the review by \citet{Nomoto13}.}
Consider, for example, a short-lived massive star.  Light elements are synthesized under hydrostatic conditions within the star starting at $\sim {\cal O}(10^4$--$10^6)$ years before the stellar core collapses.  After collapse, a hydrodynamic shock wave may develop in the stellar envelope.  Explosive nuclear burning in the shock-heated light elements synthesizes additional light and iron peak elements.  If the collapse forms a neutron star, the object initially emanates a ``wind'' of free nucleons and alpha particles.  In this wind the rapid neutron capture process (r-process) as well as a proton capture process (p-process) synthesize a family of heavy nuclei.  All of these nuclides are injected in the same galactic site---a supernova remnant that expands to become tens of light years across---in short succession. If the neutron star is left accompanied in a tight binary by another neutron star or a stellar mass black hole, gravitational radiation can induce slow inspiral and eventual coalescence of the two objects; the latter event presents a cosmically significant source of neutron capture elements. On the other hand, consider an asymptotic giant branch (AGB) star.  The slow neutron capture process (s-process) injects its own characteristic spectrum of elements beyond the iron peak. The stellar envelope's dissolution into a planetary nebula again deposits the elements locally, in a relatively tiny, light-year-size region inside a galaxy that is itself tens of thousands of light years across.  If, after a significant delay and contingent on the presence of a companion star, the white dwarf left behind undergoes thermonuclear explosion leading to a Type Ia supernova, the explosion injects elements dominated by the iron peak into a supernova remnant still only hundreds of light years across. 
  
Astrophysical nucleosynthetic sources \emph{inject their distinct admixtures of elements stochastically into spatially and temporally localized galactic sites}. Turbulent gas flow in the galaxy then gradually and stochastically homogenizes the enriched galactic gas, the interstellar medium (ISM). The homogenization process is relatively slow compared to intermittent condensation of the ISM into new populations of stars. The slow mixing implies that star-forming gas clouds are compositionally heterogeneous \citep[e.g.,][]{Scalo04}. As such, stars enclose samples of the natal gas cloud's elemental composition. The elemental composition, which can be quantified in terms of chemical element pairwise abundance ratios, can be recovered billions of years later by performing spectroscopy on the star's atmospheric absorption line system. 

Decades of theoretical nuclear astrophysical calculations belie a rich diversity of nucleosynthetic outcomes, each with its own peculiar isotopic yields. Many common as well as exotic nucleosynthetic patterns and trends have been predicted theoretically in what can be considered \emph{forward modeling} \citep[see, e.g.,][and references therein]{Nomoto13,Iliadis15,Thielemann18}.  Today, essentially-first-principles simulations of three-dimensional magnetized stellar collapse and explosion are beginning to deliver physically self-consistent predictions for isotopic yields across the periodic table, specifically including the heavy r-process isotopes---the synthesis of which is sensitive to the local conditions \citep[e.g.,][]{Halevi18}.

The goal of \emph{stellar archeology} has been to measure elemental (or isotopic) abundances in stars and interstellar and intergalactic clouds and in the solar system and match the abundance patterns to theoretical predictions \citep[see, e.g.,][]{Frebel15}.  Through this matching, the exercise of stellar archeology has aspired to reconstruct the inventory and cosmic cadence of the nucleosynthetic events that have defined the evolution of the observed universe and, among other outcomes, enabled planetary habitability. However, in an embarrassment of riches, the manifold of theoretically explored nucleosythetic outcomes, which are nonnegative vectors in as many dimensions as there are astronomically detectable isotopes, has become so large and rife with approximate parametric degeneracies that matching forward-modeled elemental abundance patterns to noisy measurements in single stars, and even in clustered groups of stars, is unable to discriminate among the specific candidate contributing nucleosythetic pathways.  

Complementing the challenging forward approach, we develop a framework for \emph{backward modeling} that aspires to \emph{reverse engineer the predominant contributing nucleosynthetic sources from vast surveys and compilations (meta surveys) of stellar chemical abundances}. Surveys such as APOGEE \citep{Hawkins15}, Gaia-ESO \citep{Mikolaitis14}, and HERMES-GALAH \citep{DeSilva15} are delivering elemental abundance data for increasingly larger samples of stars. Therefore it is becoming desirable to harness advanced data analytic algorithms to capitalize on what is potentially a rich information content uniquely buried in such comprehensive datasets. It is also clear that meta-surveys, such as the Hypatia Catalog \citep{Hinkel14} and JINAbase \citep{Abohalima17} that compile, respectively, thousands of chemical abundance measurements in the solar neighborhood, and chemical abundances in about a thousand metal poor stars in the Galaxy---all from disparate spectroscopic campaigns---present a gold mine of information. This information can then be exploited as long as one remains mindful of the nonuniform data acquisition systematics and reporting errors and biases.  

We follow in the footsteps of \citet{Ting12} who seem to have been the first to attempt purely data-driven, physical-prior-free identification of nucleosynthetic source types. For this they performed principal component analysis (PCA) and analyzed the extracted principal elemental abundance space directions. 
Our goal is also to extract archetypal elemental abundance space directions, but using a different method established in the field of \emph{blind hyperspectral unmixing}, a family of techniques used in spectroscopic identification of materials. The goal of blind hyperspectral unmixing is unsupervised discovery of pure material spectra, and of pixel-wise fractional pure material contributions, from samples of arbitrary (typically linear) superpositions of pure spectra \citep[see, e.g.,][and the references therein]{BioucasDias12,Ma14}. Since both the spectral intensities and their mixing fractions are non-negative, linear blind unmixing is a \emph{non-negative matrix factorization} (NMF) problem.  Here, we demonstrate the feasibility of simultaneous NMF and low rank completion on two real world astronomical datasets to present a second independent attempt, after \citet{Ting12}, at a physics-blind nucleosynthetic archetype discovery from stellar chemical abundance data. We illustrate the power of the method by tentatively identifying s-process, thermonuclear, and two distinct r-process-enriched nucleosynthetic archetypes in a four-way factorization of stellar abundances in the Hypatia Catalog and JINAbase.

The work is organized as follows.  Section \ref{sec:methodology} formulates the NMF and completion problem and reviews the method of solution.  Section \ref{sec:results} applies the method to the Hypatia Catalog and JINAbase and presents the recovered nucleosynthetic archetypes. Section \ref{sec:Discussion} provides a brief discussion on NMF and minimum simplex volume regularization. Section \ref{sec:conclusions} summarizes the main conclusions and motivates applying the method to the ongoing large stellar spectroscopic surveys.

\section{Methodology}
\label{sec:methodology}

\subsection{Non-negative matrix factorization}

Given a non-negative matrix $\mathbf{X}\geq 0$, we seek non-negative matrices $\mathbf{A}\geq 0$ and $\mathbf{B}\geq 0$ such that $\mathbf{A}\cdot\mathbf{B}^\text{T}\approx\mathbf{X}$. When measurements are incomplete, along with the data matrix $\mathbf{X}$, one is given an \emph{observation mask} $\mathbf{\Omega}$ such that $\Omega_{ij}=1$ when the measurement $X_{ij}$ is available and $\Omega_{ij}=0$ when it is missing.  The missing elements of $\mathbf{X}$ can be imputed under a prior on  the structure of $\mathbf{X}$.  A common prior is the assumption that $\mathbf{X}$ is a low-rank matrix perturbed by dense or sparse noise.  The low-rank property can hard-wired via the low-rank matrix factorization model that we adopt here, $\mathbf{\Omega}\circ (\mathbf{A}\cdot\mathbf{B}^\text{T})\approx \mathbf{\Omega}\circ\mathbf{X}$, where $\circ$ is the  element-wise (Hadamard) matrix product. 
Alternatively, the low-rank property can be relaxed by penalizing the loss with the nuclear norm of the reconstruction, $\min_\mathbf{Y} \|\mathbf{Y}\|_*$ such that $\mathbf{Y}\approx \mathbf{X}$, where the nuclear norm $\|\cdot\|_*$ equals the sum of the singular values of the matrix  \citep[e.g.,][]{Rennie05,Cai10,Keshavan10,Mazumder10}.   

Formal guarantees exist for correct recoverability of the missing measurements under highly idealized assumptions about the structure of $\mathbf{X}$ and $\mathbf{\Omega}$ \citep[e.g.,][]{CandesPlan10,CandesTao10}.  Among other conditions, these assumptions include the statistical independence of $\Omega_{ij}$ from $\Omega_{(i',j')\neq (i,j)}$ and  $\mathbf{X}$.  However, this independence is strongly violated in scientific data acquisition and reporting: experimenters perform measurements anticipating ``interesting'' results and shy away from reporting null results and upper limits. Simply put, published data is often ``\emph{missing-not-at-random}'' (MNAR).  If the target of reconstruction of partially observed ground truth vectors $\mathbf{x}$ under observation masks $\mathbf{\omega}$ is fully observed vectors $\mathbf{y}$, i.e., $\mathbf{\omega}\circ\mathbf{x}\approx \mathbf{\omega}\circ\mathbf{y}$, then applying a ``\emph{missing-completely-at-random}'' (MCAR) algorithm introduces biases \citep{Marlin09}. Recently, methods are being designed to reverse the MNAR bias \citep[see, e.g.,][]
{HernandezLobato14,Kim14,Yang15,Schnabel16}.  Because we anticipate the forthcoming analyses of APOGEE, Gaia-ESO, and HERMES-GALAH to involve datasets that are much less MNAR and closer to MCAR than the Hypatia Catalog or JINAbase, here we do not attempt a challenging correction for the MCAR biases.

\subsection{Chemical abundance feature vectors}

Let $x_{ij}$ denote the absolute abundance of chemical element $j$ in star $i$ (in what follows, we do not distinguish between isotopes; the elements are represented by their dominant isotopes) and let $\sigma_j$ be the uncertainty in the measurements of the abundance of the element, which we take to be the same for all stars in a compilation of abundance measurements.
We construct chemical abundance feature row vectors $\mathbf{y}_i\in \mathbb{R}_+^J$, where $J$ is the number of chemical elements measured, by rescaling the absolute abundances in the units of the measurement uncertainty
$y_{ij} \equiv x_{ij}/\sigma_j$.
We further rescale each feature vector so that its elements sum to unity. This places the vectors on the standard unit simplex:
\begin{equation}
\mathbf{z}_i\equiv \frac{\mathbf{y}_i}{\|\mathbf{y}_i\|_1} ,
\end{equation}
where, since the elements of $\mathbf{y}$ are non-negative, 
\begin{equation}
\|\mathbf{y}\|_1 = \mathbf{y}^\text{T}\cdot\mathbf{1}=\sum_{j=1}^J y_j ,
\end{equation}
and $\mathbf{1}$ denotes the vector of all ones.

Let $\mathbf{\omega}_i\in \{0,1\}^J$ denote  observation mask row vectors such that $\mathbf{\omega}_{ij}=1$ if the abundance of element $\text{X}_j$ is measured in star $i$ and $\mathbf{\omega}_{ij}=0$ if not. 
When $\mathbf{y}_i$ contains missing abundance measurements, we have no direct knowledge of $\|\mathbf{y}_i\|_1$.  Therefore we explicitly model the 
feature vector normalization with a latent factor $a_i$ 
\begin{equation}
\label{eq:scaling_defined}
\mathbf{y}_i = a_i \mathbf{z}_i, \ \ \ \| \mathbf{z}_i \|_1 = 1.
\end{equation}
We stack the row vectors $\mathbf{y}_i$, $\mathbf{z}_i$, and $\mathbf{\omega}_i$ into matrices $\mathbf{Y}\in \mathbb{R}_+^{N\times J}$, $\mathbf{Z}\in \mathbb{R}_+^{N\times J}$ where $\mathbf{Z}\cdot\mathbf{1}=\mathbf{1}$, and $\mathbf{\Omega}\in\{0,1\}^{N\times J}$, respectively, where $N$ is the number of stars in the compilation.  We further define
\begin{equation}
\mathbf{A}=\text{diag}(a_1,...,a_N).
\end{equation}
The constraint in Equation (\ref{eq:scaling_defined}) can be written as
the following optimization problem
\begin{equation}
\min_{\mathbf{A}\in \text{diag}(\mathbb{R}_+^N),\ \mathbf{Z}\in \mathbb{R}_+^{N\times J}, \ \mathbf{Z}\cdot \mathbf{1}=\mathbf{1}}  \frac{1}{2}\|\mathbf{\Omega}\circ (\mathbf{Y} - \mathbf{A}\cdot\mathbf{Z})\|_\text{F}^2,
\end{equation}
where $\|\mathbf{M}\|_\text{F}\equiv\sqrt{\sum_{ij}M_{ij}^2}$ is the Frobenius norm.  Minimization of the Frobenius loss can be seen as maximum likelihood estimation of the matrix $\mathbf{A}\cdot\mathbf{Z}$ under uncorrelated Gaussian noise $\xi(\mathbf{y})\sim {\cal N} (0,\mathbf{I})$.\footnote{The chemical abundance errors due to spectroscopic and atmospheric fitting uncertainties are almost certainly correlated between elements but the error covariances are rarely reported or discussed \citep[see, e.g.,][for an example of diagonal covariance estimation]{Hawkins15}.  The impact of error covariance on low-rank estimation will be explored in future work.  The error covariances may be fixed externally or estimated from data in a Bayesian framework.}   We select solutions of the ill-posed optimization problem by imposing a structure on the matrix $\mathbf{Z}$; specifically, we model the matrix with a product of low-rank non-negative factors.

\subsection{Low-rank factor model}

We model the normalized chemical abundance feature vectors $\mathbf{z}_i$ as convex combinations of nucleosynthetic source archetype row vectors $\mathbf{s}_r$ where $r=1,...,R$, and $R\leq J$ is the factorization rank.  The convex combinations are constructed by 
\begin{equation}
\mathbf{z}_i = \mathbf{m}_i\cdot \mathbf{S}^\text{T},
\end{equation}
where $\mathbf{m}_i\in \mathbb{R}_+^R$ is the representation row vector of convex combination coefficients $\|\mathbf{m}_i\|_1=1$ and $\mathbf{S}\in\mathbb{R}_+^{J\times R}$ is the matrix stacking the source archetype vectors.  In hyperspectral unmixing literature, the source archetype vectors are called \emph{endmembers} \citep[e.g.,][]{BioucasDias12}. Here, endmembers are synonymous with archetypes.  Stacking the convex representation vectors $\mathbf{m}_i$ into matrix $\mathbf{M}\in\mathbb{R}_+^{N\times R}$ we have
\begin{equation}
\mathbf{Z}=\mathbf{M}\cdot\mathbf{S}^\text{T}.
\end{equation}
With this, our optimization program becomes
\begin{equation}
\label{eq:objective_unpenalized}
\min_{\mathbf{A},\mathbf{M},\mathbf{S}} \frac{1}{2}\|\mathbf{\Omega}\circ (\mathbf{Y} - \mathbf{A}\cdot\mathbf{M}\cdot\mathbf{S}^\text{T})\|_\text{F}^2
\end{equation}
subject to the constraints
\begin{equation}
\label{eq:constraints}
\mathbf{A}\in \text{diag}(\mathbb{R}_+^N),\ \mathbf{M}\in \mathbb{R}_+^{N\times R},\ \mathbf{S}\in \mathbb{R}_+^{J\times R}, \ \mathbf{M}\cdot \mathbf{1}=\mathbf{1},\ \mathbf{S}^\text{T}\cdot \mathbf{1}=\mathbf{1} .
\end{equation}

To provide a geometric interpretation of the optimization program, consider the noiseless, fully observed case with rank-$R$ ground truth $\mathbf{Y}=\mathbf{A}\cdot\mathbf{M}\cdot\mathbf{S}^\text{T}$ where $\mathbf{A}=\text{diag}(\mathbf{Y}\cdot\mathbf{1})$.  The endmembers are vertices of an $(R-1)$-simplex lying within the standard unit $(K-1)$-simplex.  At optimum where the loss vanishes, the data points $\mathbf{y}_i=a_i\mathbf{m}_i\cdot\mathbf{S}^\text{T}$ lie in the interior or on the boundary of the $(R-1)$-simplex. 

\subsection{Method of solution}
\label{sec:method_of_solution}

We search for the minimum in Equation (\ref{eq:objective_unpenalized}) in alternating fashion after introducing a version of $\mathbf{Y}$ in which the missing values have been imputed via the low rank factor model
\begin{equation}
\label{eq:Y_hat_update}
\hat{\mathbf{Y}}= \mathbf{\Omega}\circ\mathbf{Y} + (\mathbf{1}-\mathbf{\Omega})\circ(\mathbf{A}\cdot\mathbf{M}\cdot\mathbf{S}^\text{T})
\end{equation}
so that
\begin{equation}
\mathbf{\Omega}\circ(\mathbf{Y}-\mathbf{A}\cdot\mathbf{M}\cdot\mathbf{S}^\text{T}) = \hat{\mathbf{Y}} -\mathbf{A}\cdot\mathbf{M}\cdot\mathbf{S}^\text{T}.
\end{equation}
Following the alternating update strategy for factorization-completion of \citet{Hastie15}, we alternate the updates $(\mathbf{M},\mathbf{S},\mathbf{A},\hat{\mathbf{Y}})_k\rightarrow(\mathbf{M}_{k+1},\mathbf{S}_k,\mathbf{A}_{k+1/2},\hat{\mathbf{Y}}_{k+1/2})\rightarrow(\mathbf{M},\mathbf{S},\mathbf{A},\hat{\mathbf{Y}})_{k+1}$ as follows:
\begin{enumerate}
\item Perform initialization by drawing the rows of  $\mathbf{M}$ and the columns of $\mathbf{S}$ randomly from the flat Dirichlet distribution. Set $k=0$.
\item $\mathbf{A}_0 = \text{diag}((\mathbf{\Omega}\circ\mathbf{Y})\cdot\mathbf{1}/\mathbf{\Omega}\cdot\mathbf{1})$.
\item $\hat{\mathbf{Y}}_0= \mathbf{\Omega}\circ\mathbf{Y} + (\mathbf{1}-\mathbf{\Omega})\circ(\mathbf{A}_0\cdot\mathbf{M}_0\cdot\mathbf{S}_0^\text{T})$.
\item \label{item:loop_goto} Update $\mathbf{M}_{k}\rightarrow \mathbf{M}_{k+1}$ by minimizing
\begin{eqnarray}
\|
\hat{\mathbf{Y}}_{k}-\mathbf{A}_{k}\cdot\mathbf{M}\cdot\mathbf{S}_{k}^\text{T}
\|_\text{F}^2.
\end{eqnarray}


\item  $\hat{\mathbf{Y}}_{k+1/2}= \mathbf{\Omega}\circ\mathbf{Y} + (\mathbf{1}-\mathbf{\Omega})\circ(\mathbf{A}_{k}\cdot\mathbf{M}_{k+1}\cdot\mathbf{S}_{k}^\text{T})$.
\item $\mathbf{A}_{k+1/2}=\text{diag}(\hat{\mathbf{Y}}_{k+1/2}\cdot\mathbf{1})$.
\item  Update $\mathbf{S}_{k}\rightarrow\mathbf{S}_{k+1}$ by minimizing
\begin{eqnarray}
\frac{1}{2} \|
\hat{\mathbf{Y}}_{k+1/2}-\mathbf{A}_{k+1/2}\cdot\mathbf{M}_{k+1}\cdot \mathbf{S}^\text{T}
\|_\text{F}^2. 
\end{eqnarray}
\item $\hat{\mathbf{Y}}_{k+1}= \mathbf{\Omega}\circ\mathbf{Y} + (\mathbf{1}-\mathbf{\Omega})\circ(\mathbf{A}_{k+1/2}\cdot\mathbf{M}_{k+1}\cdot\mathbf{S}_{k+1}^\text{T})$.
\item $\mathbf{A}_{k+1}=\text{diag}(\hat{\mathbf{Y}}_{k+1}\cdot\mathbf{1})$.
\item Test convergence of the objective and return $\mathbf{M}_{k+1}$ and $\mathbf{S}_{k+1}$ if converged.
\item $k\leftarrow k+1$, repeat from item \ref{item:loop_goto}.
\end{enumerate}
We proceed to discuss the $\mathbf{M}$ and $\mathbf{S}$ updates, as well as the initialization of $\mathbf{M}_0$, $\mathbf{S}_0$, and $\mathbf{A}_0$ in detail.

\subsubsection{The $\mathbf{M}$ Update}

Optimization with respect to $\mathbf{M}$ is performed using Nesterov's accelerated gradient descent with prox-linear update \citep{Xu13}.
The Lipschitz constant equals the largest singular value of the Hessian of the objective 
\begin{equation}
L_{\mathbf{M}} =
\|\mathbf{A}^2\|_\text{max}\ \| \mathbf{S}^\text{T}\cdot \mathbf{S}\|_2, 
\end{equation}
where $\|\cdot\|_2$ is the spectral norm equal to the largest singular value.  Let $k$ index the time step.  Define
\begin{equation}
t_0 = 1,\ \ \
t_k = \frac{1}{2} \left(1+\sqrt{1+4t_{k-1}^2}\right),\ \ \
\hat{\omega}_{k-1} = \frac{t_{k-1}-1}{t_k}.
\end{equation}
The explicit update then reads
\begin{eqnarray}
L_{\mathbf{M},k} &=& 
\|\mathbf{A}_{k}^2\|_\text{max}\ \| \mathbf{S}_{k}^\text{T}\cdot \mathbf{S}_{k} \|_2, \nonumber\\
\omega_{\mathbf{M},k} &=& \min \left(\hat{\omega}_{k},\delta_\omega \sqrt{\frac{L_{\mathbf{M},k-1}}{L_{\mathbf{M},k}}} \right),\nonumber\\
\hat{\mathbf{M}}_{k} &=& \mathbf{M}_{k} + \omega_{k} (\mathbf{M}_{k} -\mathbf{M}_{k-1}), \nonumber\\
\hat{\mathbf{G}}_{\mathbf{M},k} &=& 
\mathbf{A}_{k}\cdot(
\mathbf{A}_{k}\cdot\hat{\mathbf{M}}_{k} \cdot\mathbf{S}_{k}^\text{T}-\hat{\mathbf{Y}}_{k}) \cdot \mathbf{S}_{k},\nonumber\\
\mathbf{M}_{k+1} &=& \proj_{\mathbf{M}\cdot\mathbf{1}=\mathbf{1}}\left(\max\left(0,\hat{\mathbf{M}}_{k} - \frac{\hat{\mathbf{G}}_{\mathbf{M},k}}{L_{\mathbf{M},k}}\right)\right),
\end{eqnarray}
where `$\text{proj}$' denotes projection onto the probability simplex of the columns of $\mathbf{M}$. We perform projection with standard algorithm \citep[see][]{Chen11,Wang13}, specifically, to project $\mathbf{m}\in \mathbf{R}_+^K$ and perform the following:
\begin{enumerate}
\item Sort $\mathbf{s}$ into $\mathbf{u}$ with $u_1\geq\cdot\cdot\cdot\geq u_K$.
\item $\rho = \max\{ 1\leq j\leq K\ :\  u_j + \frac{1}{j} (1-\sum_{i=1}^j u_i) > 0  \}$.
\item $\proj(\mathbf{m})= \max(0,\,\mathbf{s}+\frac{1}{\rho}(1-\sum_{i=1}^\rho u_i)\mathbf{1})$.
\end{enumerate}

\subsubsection{The $\mathbf{S}$ Update}

Here, the Lipschitz constant is 
\begin{equation}
L_\mathbf{S} = \| \mathbf{M}^\text{T}
\cdot\mathbf{A}^2\cdot \mathbf{M}
\|_2. 
\end{equation}
The explicit update then reads
\begin{eqnarray}
L_{\mathbf{S},k} &=& \| \mathbf{M}_{k+1}^\text{T}
\cdot\mathbf{A}_{k+1/2}^2\cdot \mathbf{M}_{k+1}
\|_2, \nonumber\\
\omega_{\mathbf{S},k} &=& \min \left(\hat{\omega}_{k},\delta_\omega \sqrt{\frac{L_{\mathbf{S},k-1}}{L_{\mathbf{S},k}}} \right),\nonumber\\
\hat{\mathbf{S}}_{k} &=& \mathbf{S}_{k} + \omega_{k} (\mathbf{S}_{k} -\mathbf{S}_{k-1}), \nonumber\\
\hat{\mathbf{G}}_{\mathbf{S},k} &=& (\hat{\mathbf{S}}_{k}\cdot\mathbf{M}_{k+1}^\text{T} \cdot\mathbf{A}_{k+1/2}
-\hat{\mathbf{Y}}_{k+1/2}^\text{T})
\cdot\mathbf{A}_{k+1/2}\cdot\mathbf{M}_{k+1}, 
\nonumber\\
\mathbf{S}_{k+1} &=& \proj_{\mathbf{S}^\text{T}\cdot\mathbf{1}=\mathbf{1}}\left(\max\left(0,\hat{\mathbf{S}}_{k} - \frac{\hat{\mathbf{G}}_{\mathbf{S},k}}{L_{\mathbf{S},k}}\right)\right).
\end{eqnarray}

\subsection{Model validation}
\label{sec:tuning_validation}

Here we propose a bootstrap procedure for validating the optimal solution found by the NMF.  The procedure generates a synthetic dataset with the same endmembers and similar endmember mixing statistics, measurement noise, and observation  statistics (the structure of the random observation mask $\mathbf{\Omega}$) as the dataset being modeled.
Given a dataset ($\mathbf{X}$, $\mathbf{\Omega}$) and a proposal solution ($\mathbf{A}$, $\mathbf{M}$, $\mathbf{S}$), synthetic data can be generated as follows:
\begin{enumerate}
\item Fit a lognormal model to the distribution of the diagonal elements $a$ of $\mathbf{A}$ and draw $N$ samples from the model to obtained the resampled $\tilde{\mathbf{A}}$.
Fit a Dirichlet distribution model to the rows of $\mathbf{M}$ and draw $N$ samples from the model to obtain a resampled matrix $\tilde{\mathbf{M}}$.
\item Given $\mathbf{\Omega}$, assess element-wise (column-wise) observation probabilities and pairwise conditional observation probabilities (the probabilities of observing one element conditioned on the observation status of another element).  Perform Gibbs sampling to generate $N$ observation mask vectors consistent with the element-wise and column-wise probabilities and thus obtain a resampled observation mask $\tilde{\mathbf{\Omega}}$.
\item Set $\tilde{\mathbf{y}}_i=\tilde{a}_i \tilde{\mathbf{m}}_i\cdot\mathbf{S}^\text{T}+\xi$ where  $\xi\sim {\mathcal N}(0,\sigma_\xi \mathbf{I})$ is the measurement noise with standard deviation $\sigma_\xi$.
Stack the generated row vectors $\tilde{\mathbf{y}}_i$ into matrix $\tilde{\mathbf{Y}}\in \mathbb{R}_+^{N\times J}$.
\item For a choice of hyperparameters, perform NMF on the resampled dataset ($\tilde{\mathbf{Y}}$, $\tilde{\mathbf{\Omega}}$) to obtain a new endmember matrix $\mathbf{S}_\text{new}$.
\item  Whereas $\tilde{\mathbf{y}}_i$ are expressed in the units of measurement uncertainty and thus we expect $\sigma_\xi\approx 1$, to allow for an overall miscalibration of elemental abundance uncertainties, tune the synthetic measurement uncertainty $\sigma_\xi$ such that the average reconstruction error in Equation (\ref{eq:objective_unpenalized}) is approximately the same at optimality for real and synthetic data.  If changing $\sigma_\xi$, redo the preceding two steps.
\item Compare the fit $\mathbf{S}_\text{new}$ to the ground truth $\mathbf{S}$ by, e.g., how well can arbitrarily permuted columns of the two matrices be matched to each other under Euclidean distance.  The matching is evaluated by computing the score
\begin{equation}
\label{eq:endmember_match_score}
\ell_\mathbf{S}=
\frac{\|\mathbf{S}-\mathbf{S}_\text{new}\|_\text{F}}{
\sqrt{\mathbb{E}_\text{perm} \|\mathbf{S}-\text{perm}(\mathbf{S}_\text{new})\|_\text{F}^2}} ,
\end{equation}
where $\mathbf{S}$ and $\mathbf{S}_\text{new}$ are taken to be column-aligned and the expectation is over column permutations.
\end{enumerate}

\subsection{Model averaging}
\label{sec:model_averaging}

The NMF is non-deterministic because it requires random initialization ($\mathbf{M}_0$, $\mathbf{S}_0$) of the factor matrices and the optimization gets trapped in local optima. Therefore the factorization is repeated $T$ times, hence the resulting endmember matrices $\mathbf{S}_t$, with $t=1,...,T$, will differ. Even with perfect fidelity of the reconstruction of the ground truth endmember matrix, each $\mathbf{S}_t$ is determined only up to an arbitrary permutation of its columns.  We seek an optimal reordering of the columns of each $\mathbf{S}_t$ that maximizes the column-wise cosine similarity across all pairs ($\mathbf{S}_t$, $\mathbf{S}_{t'}$).  The reordering can be found by running the Kuhn-Munkres (`Hungarian') algorithm on the spectral embedding of the $TR\times TR$-dimensional all-endmember similarity matrix as proposed in \citet{Pachauri13}.  Column-matched $\mathbf{S}_t$ can then be directly averaged.  The averaging, of course, amplifies the distinction between the endmembers and should interpreted with caution: the cosine similarity-based matching can create dissimilar endmembers even from uncorrelated, pure-noise versions of $\mathbf{S}_t$.

\subsection{Chemical abundance datasets}

The Hypatia Catalog is a multidimensional database of high resolution stellar elemental abundances \citep{Hinkel14,Hinkel16,Hinkel17}. It is composed from over 150 individual literature sources, totaling 72 elements as measured in $\sim$6000 main-sequence stars all within 150 pc of the Sun\footnote{The full catalog and further description are available at: \url{www.hypatiacatalog.com}.}. For this work, the Hypatia data was looked at in two ways: with respect to every element abundance as measured by each literature source and as a unified dataset, per \citet{Hinkel14}. While all available stars and stellar abundances were provided in the former case, we removed some stars in the latter case when there was significant discrepancy between measurement methodologies that resulted in varying abundances. We did, however, keep those stars that likely originated from the thick disk, per the prescription by \citet{Bensby03}. In this way, we were able to determine whether those stars could be recovered from the matrix. Meaning that, as long as the measurements are not systematically erroneous, less sample purity is better for low-rank completion. For the work presented here, we restricted analysis to 37 elements in 4311 stars after excluding the  cosmogenic elements Li and Be (see the horizontal axis of the top panel of Figure \ref{fig:three}).

JINAbase is similarly structured and has a similar high level of individual element abundance incompleteness as the Hypatia Catalog, but instead of selecting for stars in the solar neighborhood, it selects for low metallicity, such that $60\%$ of the stars have $[\text{Fe}/\text{H}]<2.5$ \citep{Abohalima17}\footnote{The full catalog and further description are available at: \url{http://jinabase.pythonanywhere.com}.}. Holistically, JINAbase has a peak in the kernel density estimation metallicity distribution of [Fe/H] $= -2.75$. Its stars are mainly located in the halo, at about 90$\%$ of all unique stars, whereas the remaining fall into either the bulge, classical dwarf galaxies (Draco, Ursa Minor, etc.), and ultra-faint dwarf galaxies (Segue 1 $\&$ 2, Reticulum II, etc.). Though, there is a handful of higher metallicity thick disk stars located in the halo sample for comparison, but overall we do not expect these stars to influence our results in any statistically significant manner.  As of May 30, 2018, JINAbase contained partial measurements of the abundances of 56 elements (again, after excluding Li and Be) from 1524 distinct stars; we used all the available measurements in JINAbase.

\begin{figure*}
\begin{center}
\includegraphics[width=\textwidth]{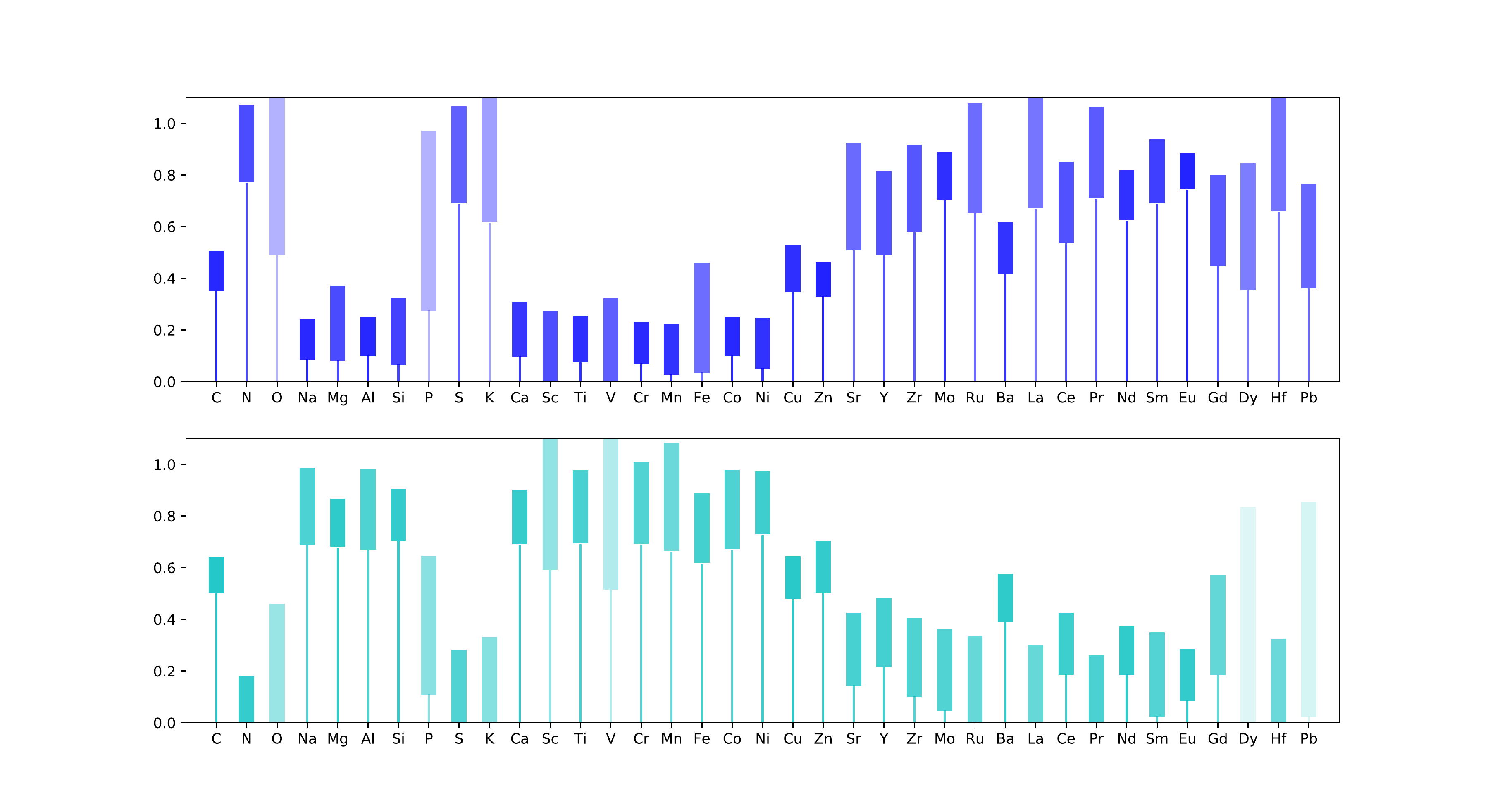}\vspace{1cm}
\includegraphics[width=\textwidth]{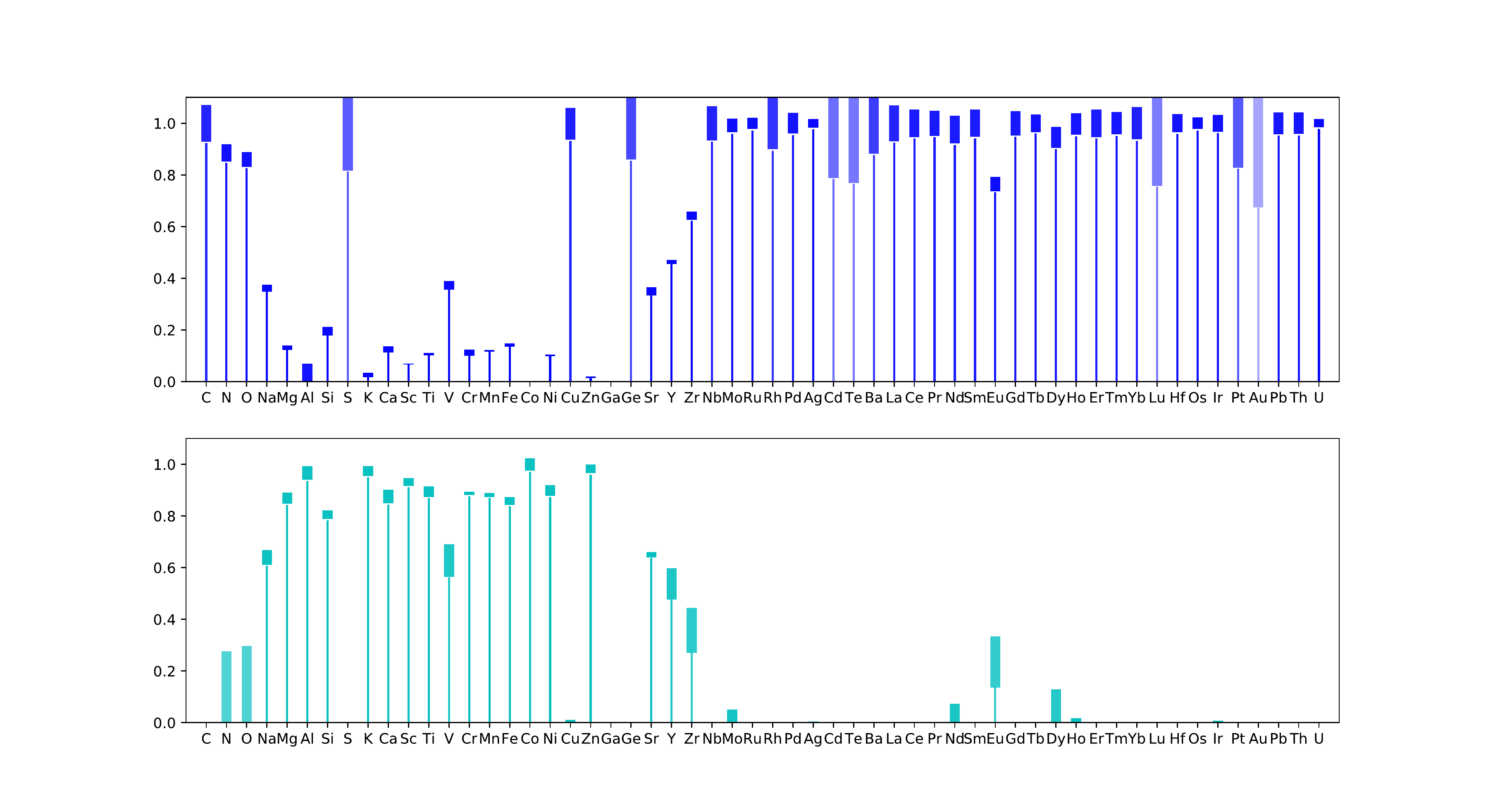}
\end{center}
\caption{Rank $R=2$ non-negative matrix factorization endmembers for the Hypatia Catalog (top two panels) and JINAbase (bottom two panels).
The column height for each individual element is computed via the model averaging explained in Section \ref{sec:model_averaging} and is normalized to sum to unity across the endmembers. The block error bars show $\pm1\sigma$ model averaging dispersion.
\label{fig:two}}
\end{figure*}

\begin{figure*}
\begin{center}
\includegraphics[width=\textwidth]{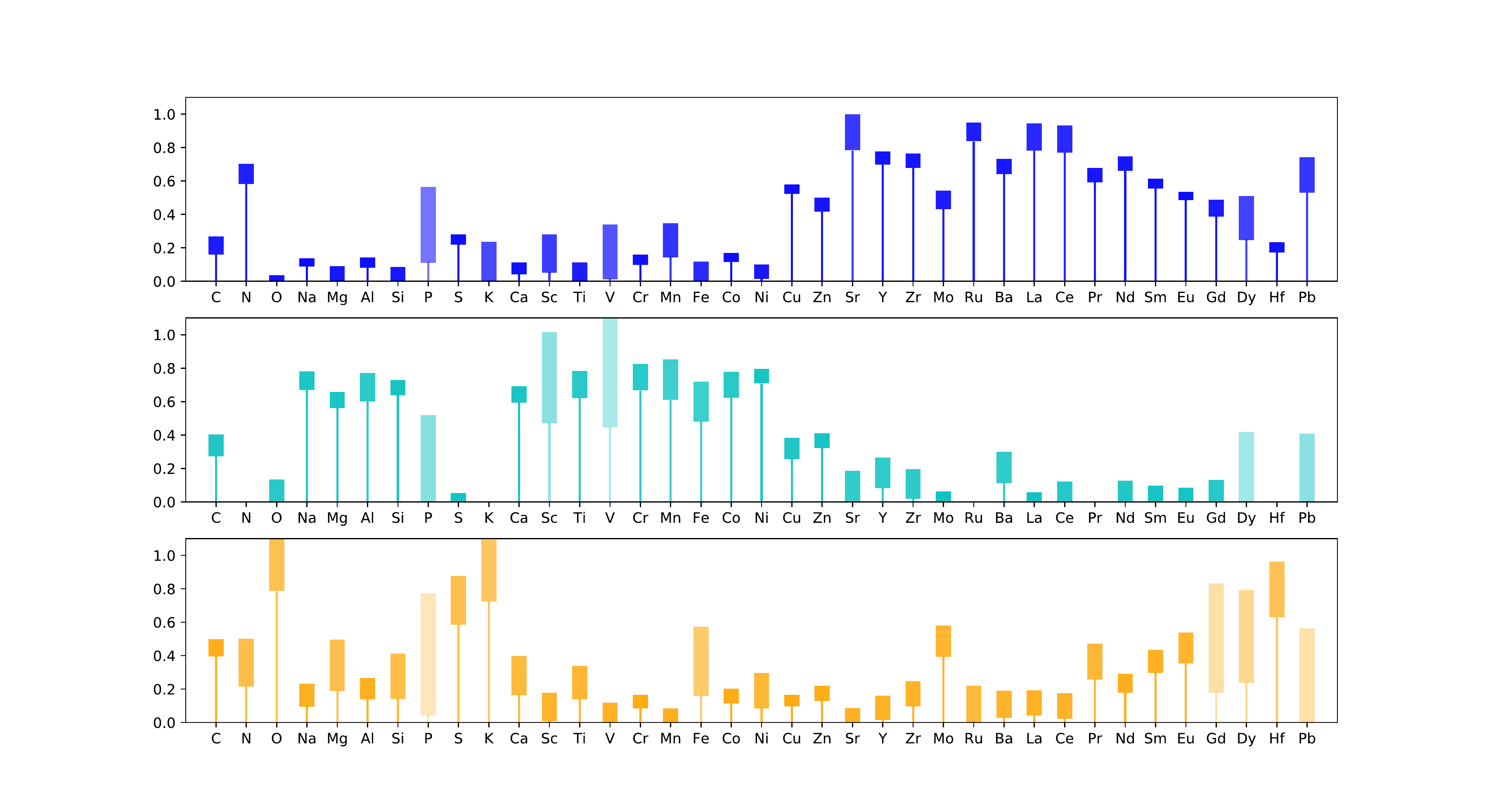}\vspace{1cm}
\includegraphics[width=\textwidth]{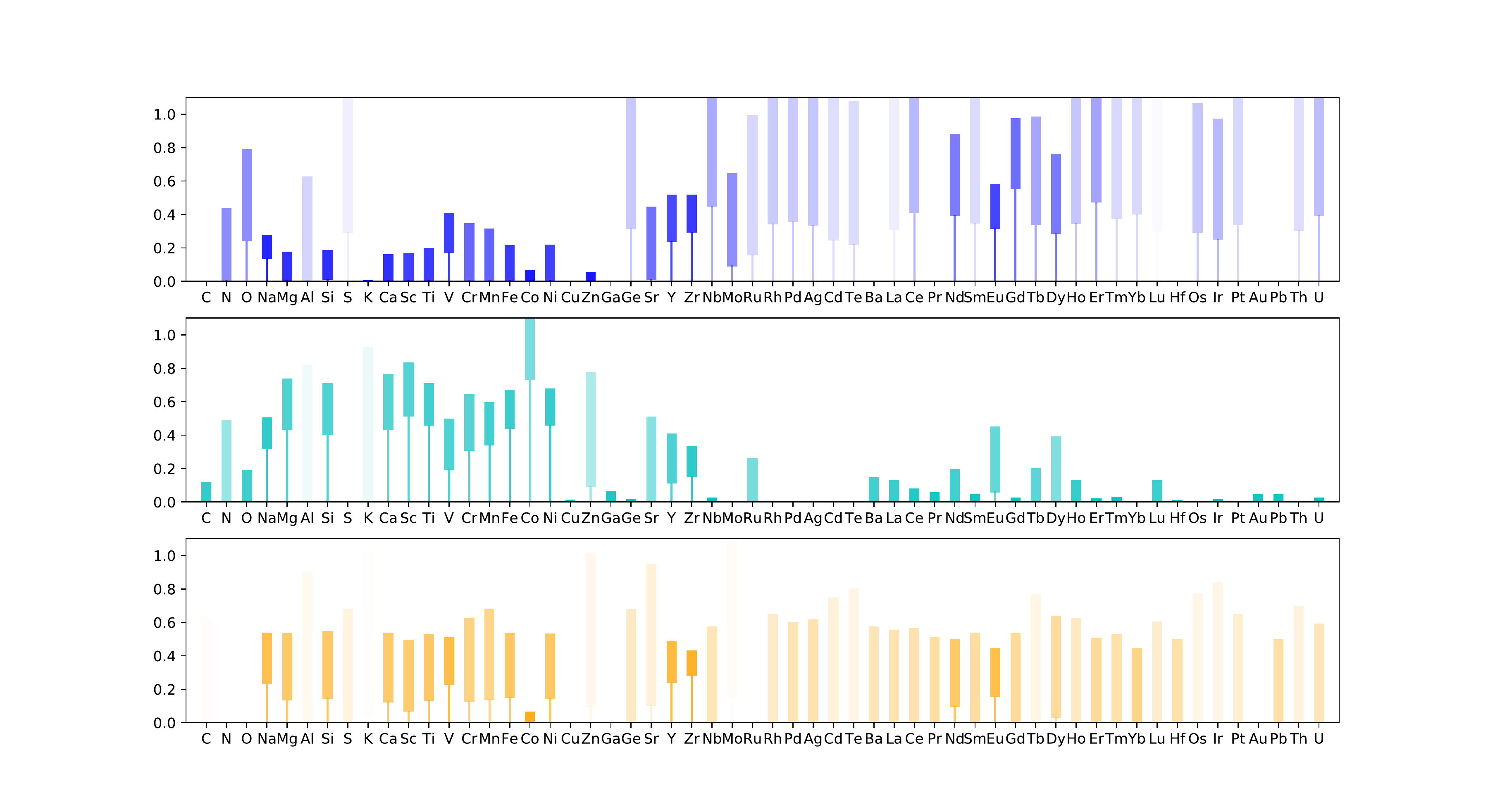}
\end{center}
\caption{The same as Figure \ref{fig:two}, but for $R=3$ factorization of the Hypatia Catalog (top three panels) and JINAbase (bottom three panels).
\label{fig:three}}
\end{figure*}

\begin{figure*}
\begin{center}
\includegraphics[width=\textwidth]{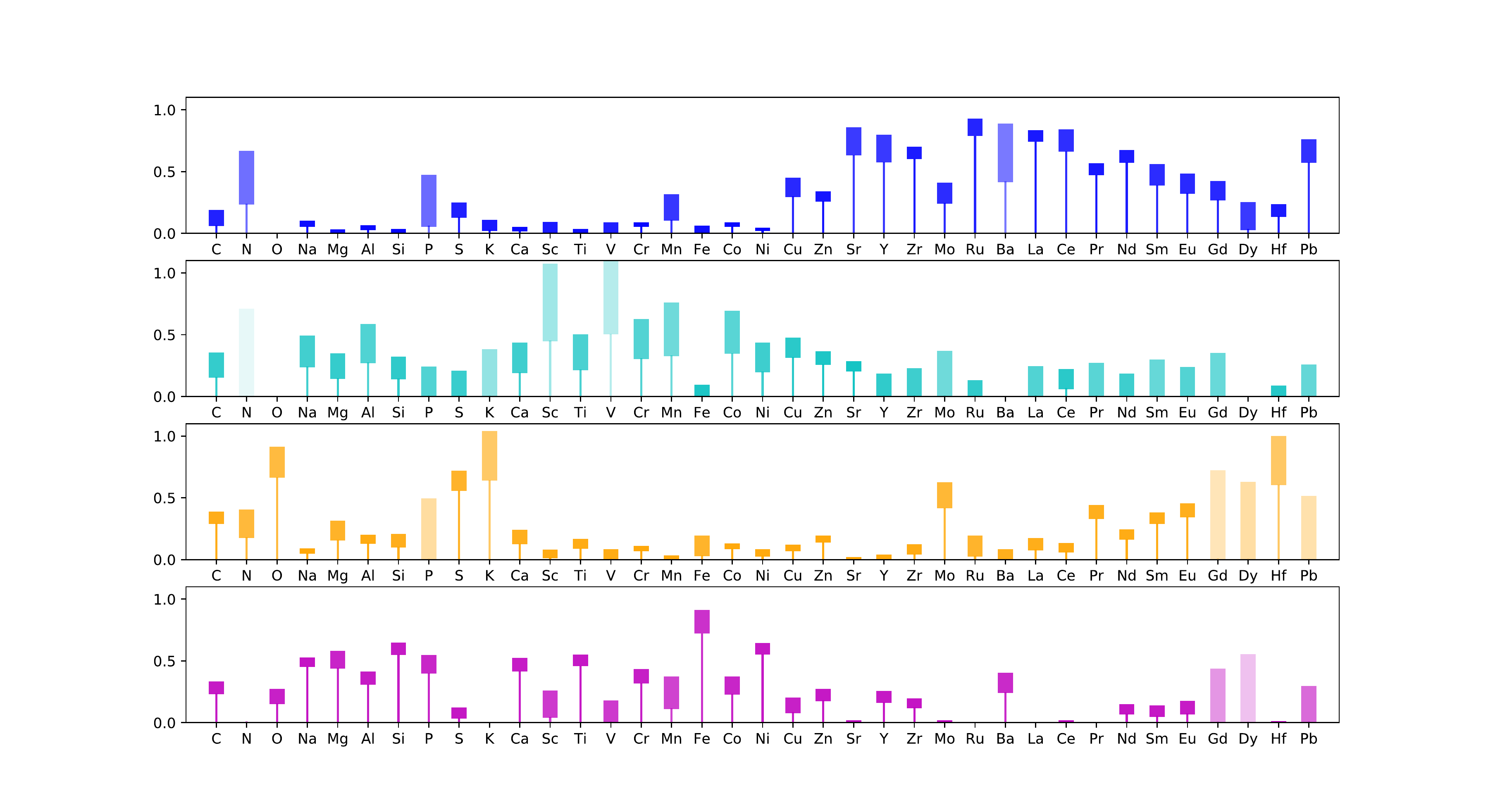}
\end{center}
\caption{The same as Figure \ref{fig:two}, but for $R=4$ factorization of the Hypatia Catalog.
\label{fig:four}}
\end{figure*}

\section{Results}
\label{sec:results}

To validate the success of the simultaneous NMF and completion, for each factorization rank $R\geq 2$, we perform an initial fit to the Hypatia dataset (and JINAbase for all that follows) to obtain an endmember matrix $\mathbf{S}$. Then, as described in Section \ref{sec:tuning_validation}, we generated synthetic realizations of Hypatia-like datasets.  Matching the average real and synthetic reconstruction errors in Equation (\ref{eq:objective_unpenalized}) required recalibrating the measurement noise to $\sigma_\xi\approx 0.3$.  We used the score in Equation (\ref{eq:endmember_match_score}) to quantify how well the simultaneous NMF and completion could recover the input $\mathbf{S}$.  We take the successful recoveries to be $\ell_\mathbf{S}\leq 0.2$.  Recoveries can be unsuccessful due to entrapment in local minima of the NMF objective and can be screened out by selecting the factorizations with the lowest objectives. The recovery success rate is $100\%$ for ranks $R\leq 4$.  For $5\leq R\leq 11$ the success rate is below one hundred percent and diminishing with increasing rank.  In what follows, we report the $R=2$ and $R=3$ factorizations for both datasets as well as the $R=4$ factorization for the Hypatia Catalog.  

In Figure \ref{fig:two} we present an average, computed as described in Section \ref{sec:model_averaging}, of $100$ NMFs with $R=2$ endmembers, which are our nucleosynthetic archetypes.   Before we hazard an astrophysical interpretation of the archetypes, we remind the reader that the analysis up to this point has been completely astrophysically agnostic. The extracted archetypes do not ``know'' about astrophysically anticipated abundance patterns, unless, perhaps, some astrophysical foreknowledge has snuck in through the back door of the MNAR statistics of the observation pattern $\mathbf{\Omega}$ that reflects various astrophysical biases.  

We caution that the actual number of nucleosynthetic source archetypes we expect to be represented in either the solar neighborhood (Hypatia Catalog) or among metal poor stars (JINAbase) is almost certainly larger than two or three, thus if restricting to a very small number archetypes, partial blending of the archetypes is inevitable.  We choose to present such a small number of archetypes because the recovery success rate decreases with increasing archetype number $R$.  We further caution that \emph{any nucleosynthetic source that is well mixed}, in the sense that it contributes in equal proportion to all the stars in the dataset---such as due to turbulent mixing in the course of infall from the circumgalactic medium if the chemical pollutants are first deposited outside of the Galactic gaseous disk---will remain unidentified.

Figure \ref{fig:two} shows that the $R=2$ archetype factorizations of the Hypatia Catalog and JINAbase are similar. The first archetype associates C, N, and O and the neutron capture element Sr and beyond.  In JINAbase, there is a monotonic increase of association in the first archetype from Sr through Nb. The first archetype clearly represents \emph{hydrostatic} nucleosynthesis, albeit with some outliers that we discuss individually. The element S is an apparent outlier but its abundance is measured only in 2 stars ($0.1\%$) in JINAbase, compared to $15\%$ of stars in the Hypatia Catalog.  Similarly, the element P is an outlier in the Hypatia Catalog factorization but its abundance is measured in only 20 stars there ($0.5\%$).  The element K is an outlier in the Hypatia Catalog with measurements in $7\%$ of the stars.  In the JINAbase factorization, the element Cu, measured in $3\%$ of the stars, is an outlier, as is the element Ge measured in only $0.3\%$ of the stars.

The second archetype associates the light elements Na through Si, as well as the explosively synthesized elements Ca (K in JINAbase) through Ni (Zn in JINAbase).  There is also a clear tail of diminishing association of the ``light'' r-process elements starting at the edge of the iron peak (Ni in the Hypatia Catalog, Zn in JINAbase) and extending through Mo. This tail could indicate the photon ($\gamma$) or neutrino ($\nu$) p-process. 
The second archetype therefore clearly associates elements produced in \emph{explosive} nucleosynthesis, albeit with the same outliers as in the first archetype.  The heavy neutron capture elements exhibit noisy association in the Hypatia Catalog, and among them, the association is the strongest in Ba, Ce, Nd, Eu, and Gd.  In JINAbase, only Eu is significantly associated with the second archetype, a hint of the heavy r-process.
It is interesting that the lightest element strongly associated with the explosive archetype is Na.

Turning to the $R=3$ factorization in Figure \ref{fig:three}, differences begin to appear between the Hypatia Catalog and JINAbase factorizations.  In the Hypatia factorization, the first archetype remains clearly associated with \emph{AGB star} nucleosynthesis; it associates C and N with the s-process elements, most strongly in Sr through Ce.  Interestingly, from Ce through Hf, the association with the first archetype steadily declines, which suggests an alternate neutron capture source, one that is clear in the third archetype of the Hypatia Catalog factorization.  Indeed, the third archetype exhibits some of the associations of \emph{core-collapse-type} nucleosynthesis: C, O, Mg, Si, S, K, Ca, Ti, Fe, Ni, Mo, as well as a potential heavy r-process peak in Pr through Hf.  The second archetype in the Hypatia Catalog factorization strongly associates Ca through Ni, as well as the light elements Na through Si, and no significant neutron capture elements, consistent with \emph{thermonuclear} nucleosynthesis in Type Ia supernovae.  If the three archetypes of the $R=3$ factorization of the Hypatia Catalog indeed discern AGB, thermonuclear nucleosynthesis, and core collapse, then this places the origin of Na in the context of thermonuclear burning.

The $R=3$ factorization of JINAbase, however, does not show this three-way split.  Its first two archetypes strongly resemble the $R=2$ JINAbase factorization that we have tentatively interpreted as hydrostatic and thermonuclear nucleosynthesis. The second archetype now exhibits a weak association with the heavy r-process elements, primarily Eu.  The third archetype of the JINAbase factorization associates all the elements apart from C, N, and O approximately equally, suggesting this archetype is not well differentiated.  We speculate that in $R=3$ JINAbase factorization, the thermonuclear and core collapse archetypes remain fused in the single, second archetype.  The difficulty of splitting the two may stem from the stars in JINAbase being selected to be low in Fe.

Finally, in Figure \ref{fig:four}, we show the $R=4$ factorization of the Hypatia Catalog. The first archetype is the now familiar hydrostatic, AGB star association.  The second archetype again associates the light elements starting with Na and extending through Mn, skipping Fe, and then again from Co through Sr.  Similarly, as in the $R=3$ factorization, third archetype most strongly associates O, S, K, as well as Mo, suggesting explosive nucleosynthesis, as well as the heavy neutron capture elements Pr through Hf.  The new, fourth archetype associates the light and iron peak elements C through Ni, but now with a prominent \emph{even over odd preference} from Mg through Zn, suggesting an $\alpha$ chain.  There is also an association with Ba and a  weak association with the light (Y, Zr) and heavy (Nd through Dy; Pb) r-process.

\subsection{Comparison with \citet{Ting12}}

\citet{Ting12} performed PCA on stellar elemental abundances and extracted four principal components. They attempted to interpret each principal component in terms of a distinct nucleosynthetic mechanism.  They associate the production of $\alpha$ and $r$-process elements, hinting at core collapse supernovae as a potential r-process site.  However, they only detected this association in their low-metallicity ($-3.5 \leq$ [Fe/H] $\leq -2$) but not the high-metallicity ([Fe/H] $\geq -1$) sample.  A majority of Hypatia Catalog stars fall in this (comparatively) high-metallicity regime with $-0.2 \leq$ [Fe/H] $\leq 0.5$, though the Hypatia Catalog contains does contain a handful of thin-disk stars with [Fe/H] $\textless -1.0$ dex. In both their low and high metallicity samples, they find one principal component dominated by Cr and Mn, with little contribution from the other elements, whereas we find Cr and Mn to be prominent only when the other Fe-peak elements are strong. \par
Lastly, \citet{Ting12} also allude to two distinct core collapse sources: one which produces predominately $\alpha$ elements and another which produces both $\alpha$ and iron elements including the heavy tail of the iron peak.

\section{Discussion}
\label{sec:Discussion}

Approximate NMF solutions tend to return sparse factors, that is, a fraction of the entries of $\mathbf{A}$ and $\mathbf{B}$ is zero \citep[see, e.g.,][and the references therein]{Vandaele16,Gillis17}. Factor sparsity promotes interpretability in NMF applications in which each observation is expected to contain only a small subset of contributing materials and/or each material has telltale emission signatures at only a small number of specific wavelengths.  In contrast, the mixing fraction vectors of astrophysical nucleosynthetic source archetypes in an evolved galactic environment, the rows of both $\mathbf{A}$ (the fractional contributions of nucleosynthetic sources) and $\mathbf{B}$ (the fractional chemical abundances in distinct nucleosynthetic sources), are expected to be dense in the sense that most of its components are non-zero. Because of this density, blind unmixing of elemental abundance archetypes in, say, the Milky Way disk, must rely on fluctuations on the top of an otherwise well-mixed baseline.\footnote{An alternative strategy is to design astronomical surveys so as to search for  specific stars that seem as \emph{pure} as possible in terms of the number of distinct nucleosynthetic source archetypes required to explain them \citep[e.g.,][]{Hansen15}. Samples known to contain such pure samples are called \emph{separable} in the hyperspectral unmixing literature.}  

The source archetype vectors, the columns of $\mathbf{B}$, may be dense or sparse, but interpretable archetypes should be dense provided that: (1) the archetype dictionary is parsimonious and \emph{not} overcomplete, i.e., the archetype number is as small as possible, and (2) the decomposition of stellar abundance records into archetypes is subject to Occam's razor, i.e., the representation space disfavors instances even purer in specific archetypes than the training data.  Since the representation space is a \emph{convex hull} of the archetype vectors, Occam's razor can be construed as ``minimization'' of this convex hull.  

If the rows of $\mathbf{X}$ and $\mathbf{A}$ and the columns of $\mathbf{B}$ are normalized to lie on the unit simplex (which is the set of nonnegative vectors with unit $\ell_1$ norm), then Occam's razor translates into a minimization of a function representative of the volume of the convex hull. Minimum-simplex-volume-regularized non-negative matrix factorization is a state-of-the-art technique for blind hyperspectral unmixing \citep[][]{Miao07,Chan09,Fu16}.  Recently, \citet{Javadi17} proposed a closely related method framed in the language of archetypal analysis.  We extensively experimented with minimum simplex volume regularization but found that on the Hypatia Catalog and synthetic Hypatia-like data the regularization did not perceptibly improve interpretability of the NMF. Therefore, in the interest of simplicity, here we have presented only the method and results of unregularized NMF and plan to revisit minimum simplex volume regularized NMF in future work.

\section{Conclusions}
\label{sec:conclusions}

We performed joint non-negative matrix factorization and low-rank completion on incomplete 37 element chemical abundance measurements in 4311 stars of the Hypatia Catalog and incomplete 56 element chemical abundance measurements in 1524 stars of JINAbase. The factorizations ($R$ = 2, 3, 4) provides nucleosynthetic archetypes from which, within the bounds of measurement uncertainties, the abundances of the observed stars can be approximated up to an overall normalization factor as convex combinations. We tentatively discern astrophysically interpretable archetypes: AGB stars with s-process, thermonuclear supernovae, and non-thermonuclear explosive nucleosynthesis. The physics-blind nucleosynthetic archetype discovery procedure presented here is ideally suited for the new, larger, and more complete and homogeneous chemical abundance samples from APOGEE, Gaia-ESO, and HERMES-GALAH.

\section*{Acknowledgements}

We acknowledge conversations with the astrophysicists S.\ Couch, K.\ Hawkins, I.\ Ramirez, C.\ Sneden, Y.-S.\ Ting, and J.\ C.\ Wheeler, as well as the applied mathematicians X.\ Fu, K.\ Ma, B.\ Mishra, N.\ D.\ Sidiropoulos, and Y.\ Xu that have helped inform this work.  
This study was supported by the NSF grant AST-1413501 and was performed in part at the Aspen Center for Physics, which was supported by National Science Foundation grant PHY-1066293.
NRH was supported by the Vanderbilt Office of the Provost through the Vanderbilt Initiative in Data-intensive Astrophysics (VIDA) fellowship.
The research shown here acknowledges use of the Hypatia Catalog Database, an online compilation of stellar abundance data as described in \citet{Hinkel14}, which was supported by NASA's Nexus for Solar System Science (NExSS) research coordination network and VIDA.

\markboth{Bibliography}{Bibliography}

\bibliographystyle{mnras}
\bibliography{nucleo.bib}

\end{document}